\documentclass[journal]{IEEEtran}

\usepackage{amsmath}
\usepackage{cite}
\usepackage{color}
\usepackage{mdwtab}
\usepackage{subfigure}
\usepackage{placeins}
\usepackage{psfrag}
\usepackage{graphicx}
\usepackage{epsfig}
\usepackage{epstopdf}
\usepackage[latin1]{inputenc}
\usepackage{amssymb}
\usepackage{amsthm}
\usepackage{comment}
\usepackage{setspace}

\theoremstyle{plain}
\newtheorem{theorem}{Theorem}

\theoremstyle{remark}
\newtheorem{lemma}{Lemma}

\IEEEoverridecommandlockouts

\begin{document}

\title{Resolution-Based Content Discovery in Network of Caches: Is the Control Traffic an Issue?} 

\author{Bita~Azimdoost,
        Cedric~Westphal,~\IEEEmembership{Senior Member,~IEEE,}\\
		and Hamid~R.~Sadjadpour,~\IEEEmembership{Senior Member,~IEEE,} 
   
\thanks{B. Azimdoost and H. R. Sadjadpour are with the Department of Electrical Engineering,
University of California, Santa Cruz, 1156 High Street, Santa Cruz,
CA 95064, USA (e-mail:\{bazimdoost, hamid\}@soe.ucsc.edu)}
\thanks{Bita Azimdoost was with Huawei Innovation Center, Santa Clara, CA 95050, USA, as an intern while working on this paper.}
\thanks{Cedric Westphal is with the Department of Computer Engineering,
University of California, Santa Cruz, 1156 High Street, Santa Cruz,
CA 95064, USA and Huawei Innovation Center, Santa Clara, CA 95050, USA (e-mail:cedric.westphal@huawei.com)}}

\maketitle

\begin{abstract}
As networking attempts to cleanly separate the control plane and forwarding plane abstractions, it also defines a clear interface between these two layers. An underlying network state is represented as a view to act upon in the control plane. We are interested in studying some fundamental properties of this interface, both in a general framework, and in the specific case of content routing. We try to evaluate the traffic between the two planes based on allowing a minimum level of acceptable distortion in the network state representation in the control plane.

We apply our framework to content distribution, and see how we can compute the overhead of maintaining the location of content in the control plane. This is of importance to evaluate resolution-based content discovery in content-oriented network architectures: we identify scenarios where the cost of updating the control plane for content routing overwhelms the benefit of fetching the nearest copy. We also show how to minimize the cost of this overhead when associating costs to peering traffic and to internal traffic for network of caches.
\end{abstract}

\section{Introduction}
A communication network can be abstracted into two (logical) layers, namely, a control plane carrying signaling and administrative traffic, and a data forwarding plane carrying the user data traffic. In many applications, for the network to function properly, the control plane must have some knowledge about the forwarding plane in order to create a view of the underlying network. The underlying network will be in an operating state which is reported by a protocol to the control/management layer. For example, in a network of caches, the data plane contains caches keeping the data traffic, e.g. video or audio files, which are requested and used by the users, and the information regarding the items kept in each cache reported to the control layer forms the control traffic. 

However, as the networks have grown in size and complexity, as end nodes, content and virtual machines move about, it will become more difficult for the control layer to have an accurate view of the forwarding plane. Consider the example of finding a service or a piece of content. Current protocols attempt to resolve a content request to the nearest copy of the object by using DNS or redirecting HTTP requests. Further proposals suggest to share content location information in between content delivery networks (CDNs), or even to build content routing within the architecture. In all cases, this implicitly entails that the mechanism responsible to route to the content has to be dynamically updated with the content location. Meta-information from the forwarding plane needs to be delivered to the control plane. This raises the question: {\em how much?} In other words, depending on the size of the domain being controlled, of the underlying state space, of the dynamics of the evolution of the state in the forwarding plane, what stream of data is required to keep the control plane up to date?

We consider the issue of maintaining a consistent view of the underlying state at the control layer, and develop an abstracted mechanism, which can be applied to a wide range of scenarios. We assume the underlying state as an evolving random  process, and calculate the rate that this process would create to keep the representation of this state up-to-date in the control plane. This provides a lower bound on the overhead bandwidth required for the control plane to have an accurate view of the forwarding plane\footnote{There  exists other overhead that we have not discussed in this work. We believe that addressing all the overhead of data/control plane interface in one paper may not be possible, since there might be several sources for them. However, if one thinks of certain sources of overhead, like the overhead of setting up a secure connection between the forward and control planes, then the actual protocol overhead would be proportional to the information theoretic overhead, at least if the rate of update is high enough. In which case we provide a good idea of how the whole protocol overhead will trend.}.

We then illustrate the power of our model by focusing on the specific case of locating content in a resolution-based content-oriented network. Enabling content routing has attracted a lot of attention recently, and thus we are able to shed some light on its feasibility.  In this case, the underlying state depends on the size and number of caches, on the request for content process and on the caching policy. We  apply our framework to derive the bandwidth needed to accurately locate a specific piece of content. We observe that there is a  trade-off for keeping an up-to-date view of the network at the cost of  significant  bandwidth utilization, versus the gain achieved by fetching the nearest copy of the content. We consider a simple scenario to illustrate this trade-off.

Our contribution is as follows:
\begin{itemize}
\item We present a framework to quantify the minimal amount of information required to keep a (logical) control plane aware of the state of the forwarding plane. We believe this framework to be useful in many distributed systems contexts.
\item We apply our framework to the specific case of locating content, and see how content location is affected by the availability of caches, the caching policy and the content popularity. We can thus apply our results to some of the content-oriented architectures and observe that cached copies would go ignored for a large swath of the content set.
\item We see how our framework allows to define some optimal policies with respect to the contents that should be cached for an operator-driven content distribution system. While it is not surprising that very unpopular contents should not be cached, we can actually compute a penalty for doing so under our model.
\end{itemize}

We quickly note that our framework does not debate the merit of centralized vs distributed, as the control layer we consider could be either. For a routing example, our model would provide a lower-bound estimate of the bandwidth for, say OpenFlow to update a centralized SDN controller, or for a BGP-like mechanism to update distributed routing instances.

Our results are theoretic in nature, and provide a lower bound on the overhead. We hope they will provide a practical guideline for protocol designers to optimize the protocols which synchronize the network state and the control plane.

The rest of the paper is organized as follows. After going over some related work in section \ref{sec:related}, we introduce our framework to model the protocol overhead in section \ref{sec:framework}, and then study the content location in the network of caches in section \ref{sec:scenarios}. The derived model is used to study a simple caching network as well. We show the power of the model in the protocol design by computing the cost of content routing in Section~\ref{sec:costanalysis} and suggesting a cache management policy. Finally, section \ref{sec:conclusion} concludes the paper and describes some possible future work.

\section{Related Work}
\label{sec:related}

As SDN makes the separation explicit between the control and forwarding layers, it begs the question of how these layers interact. This interaction has been pointed out as one of the bottlenecks of OpenFlow~\cite{McKeown2008OpenFlow}, and several papers have been trying to optimize the performance of the traffic going from one layer to the other. For instance, \cite{Tootoonchian2012Controller} optimizes the controller to support more traffic, while \cite{Curtis2011Devo} or \cite{Yu2010Difane} attempt to make the control layer more distributed and thus reduce the amount of interaction  between the switches and the control layer. There has been no attempt to model the interaction between the control and forwarding layers to our knowledge.

Studying the gap between the state of the system and the view of the controller, \cite{Levin2012Logically} focuses on the relationship between performance and state consistency, and \cite{Bari2013Dynamic} studies similar relationship in multiple controller systems. This underlines the need for the view at the control layer to be representing the network state with as little distortion as possible.

The forwarding plane in a network usually consists of a state machine which is changing because of different network characteristics. The control plane needs to obtain adequate information about the underlying states so that the network can perform within a satisfactory range of distortion. The first theoretical study of this information was conducted by Gallager in \cite{Gallager1976Basic}. This work utilizes the rate distortion theory to calculate the bounds on the information required to show some characteristics such as the start time and the length of the messages.

The link states (validity of a link) and the geographic location and velocity of each node in a mobile wireless network are some examples of such state, which have been studied in \cite{Wang2008Link} and \cite{Wang2012Cost}, respectively.   An information-theoretic framework to model the relationship between network information and network performance, and the minimum quantity of information required for a given network performance was derived in \cite{Hong2009Impact}.

One impetus to study the relationship between the control layer and the network layer comes from the increased network state complexity from trying to route directly to content. Request-routing mechanisms have been in place for a while\cite{Barbir2003Requestrouting} and proposals~\cite{Davie2012Framework} have been suggested to share information between different CDNs, in essence enabling the control planes of two domains to interact (our framework applies to this situation). And many architectures have been proposed that are oriented around content\cite{Gritter2001Architecture,Koponen2007Dataoriented,Jacobson2009Networking,Zhang2010Named,Trossen2011PURSUIT,Ahlgren2012Survey} and some have raised concerns about the scalability of properly identifying the location of up to $10^{15}$ pieces of content\cite{Ghodsi2011InformationCentric}. Our model presents a mathematical foundation to study the pros and cons of such architectures.

The cache management problem in the networks has been studied in several contexts. \cite{Tang2008BenefitBased} presents a centralized approximation algorithm to solve the cache placement problem for minimizing the total data access cost in ad hoc networks. \cite{Bhattacharjee1998Self} proposes a replication algorithm that lets nodes autonomously decide on caching the information, and \cite{Azimdoost2015Optimal} determines whether/where to keep a copy of a content such that the overall cost of content delivery is minimized and show that such optimized content delivery significantly reduces the cost of content distribution and improves quality of service.

Some cooperative cache management algorithms are developed in \cite{Borst2010Distributed} which tries to maximize the traffic volume served from cache and minimize the bandwidth cost in content distribution networks. \cite{Sourlas2012Autonomic} proposes some online cache management algorithms for Information Centric Networks (ICNs) where all the contents are available by caching in the network instead of a server or original publisher.
\cite{Chai2012Cache} investigates if caching only in a subset of node(s) along the content delivery path in ICNs can achieve better performance in terms of cache and server hit rates. These works define a specific cost in the network and try to determine the locations and the number of copies of the contents in the network such that the defined cost is minimized. Finally \cite{Azimdoost2013Throughput} and \cite{Azimdoost2016Fundamental} analytically prove that on-path content discovery has the same asymptotic capacity as finding the nearest copy in these networks. 

To the best of our knowledge, there is no work considering the protocol overhead in such systems. In this work, we model the protocol overhead, then use that model to compute a general cost for data retrieval (including the protocol overhead). We also investigate whether allowing more copies of the contents cached in the network reduces the total cost. One related work on this topic  is \cite{Cho2012WAVE} which proposes a content caching scheme, in which the number of chunks (fragments) to be cached in each storage is adjusted based on the popularity of the content. In this work, each upstream node recommends the number of chunks to be cached in the downstream node according to the number of requests.

\section{Protocol Overhead Model}
\label{sec:framework}

In this section we turn our attention to the mechanism to synchronize the view at the control layer with the underlying network state, and introduce a framework to quantify the minimal amount of required transferred information. 

Assume that $S_X(t)$ describes the state of random process $X$ in a network at time $t$. In order to update the control plane's information about the states of $X$ in the network, the forwarding plane must send update packets regarding those states to the control plane whenever some change occurs. Let $\hat{S}_X(t)$ denote the control plane's perceived state of $X$ at time $t$. It is obvious that no change in $\hat{S}_X$ will happen before $S_X$ changes, and if $S_X$ changes, the control plane may or may not be notified of that change. Therefore,  there are some instances of time where $\hat{S}_X\ne S_X$.

\begin{figure}[http]
    \center
      \includegraphics[scale=0.65,angle=0]{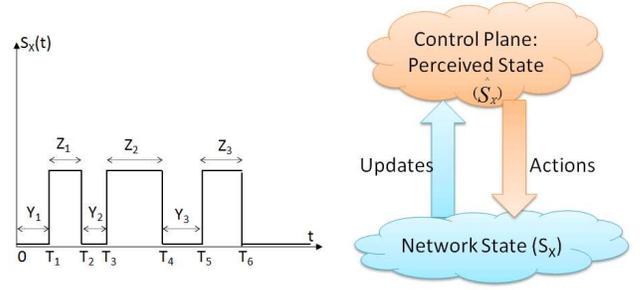}\\
      \caption{\textit{Time diagram of the state of binary random process $X$ at time $t$ ($S_X(t)$).}}
    \label{fig:statetime}
\end{figure}

In this paper, we consider, systems and applications in which the state can have two values $'0'$ and $'1'$.\footnote{Note that this Boolean case is just an example to illustrate the method, and can be generalized to other possible values. For instance, to measure the congestion on a link, one could quantize the link congestion into bins (say bins $b_1$ to $b_{10}$ for normalized link utilization between $0$ to $0.1$, $0.1$ to $0.2$,..., $0.9$ to $1$) and map the link utilization to a $0/1$ variable such that $b_h=1$ if the current link utilization is in ($(h-1)/10$, $h/10$) and $0$ otherwise. Obviously using this quantization method the $b_h$ variables would not be independent and only one of them can be $'1'$ at each instant of time. As other way to solve such problem, one can model the changes in the quantized levels as a binary variable. Since the values of the congestion levels change smoothly and there is not any kind of discontinuity in the congestion levels, one can expect going one level up or down in case of any changes. Using this method, one needs to have new distortion definitions. Due to the lack of space we leave it as future work to study other state distributions, where other distortion functions would apply.} For instance, a link can be up or down; or a piece of content can be present at a node, or not.  Figure \ref{fig:statetime} illustrates the time diagram of state changes of such binary random process which is the state of the forwarding plane in the network being announced to the control plane.

Let $\{Y_m\}_{m=1}^{\infty}$ and $\{Z_m\}_{m=1}^{\infty}$ denote the sequences of $'0'$s and $'1'$s time durations of $S_X(t)$ respectively, and $\{T_m\}_{m=1}^{\infty}$ denote the times of changes. We consider large distributed systems, where the input is driven by a large population of users (smaller systems offer no difficulty in tracking in the control plane what is happening in the data plane). It is a well known result that the aggregated process resulting from a large population of uncoordinated users will converge to a Poisson process (chapter 3.6 \cite{durrett2010probability}), and therefore the events in the future are independent of the events in the past and
depend only on the current state. Thus we assume with no loss of generality that $Y_m$ is an independent and identically distributed (i.i.d) sequence with probability density function (pdf) $f_Y(y)$ and mean $\theta_X$, and $Z_m$ is another i.i.d. sequence with pdf $f_Z(z)$ and mean $\tau_X$. We also assume that any two $Y_m$ and $Z_m$ are mutually independent.

 $S_X$ and $\hat{S}_X$ may differ in two cases resulting in two types of distortion; first, when the state of $X$ is changed from $'0'$ to $'1'$ (change type I) but the control plane is not notified ($\hat{S}_X=0,\ S_X=1$); second, when the state of $X$ is changed from $'1'$ to $'0'$ (change type II) and the control plane still has the old information about it ($\hat{S}_X=1,\ S_X=0$). Let $D_1$ and $D_2$ denote the probability of the distortions corresponding to changes type I and II, respectively. Here we calculate the minimum rate at which the underlying plane has to update the state of $X$ so that the mentioned distortion probabilities are less than some values $\epsilon_1$ for the first type, and $\epsilon_2$ for the second type, respectively. $\epsilon_1$ and $\epsilon_2$ can be viewed as probability of false negative and false positive alarms at the controller. 

We make an additional assumption that the delay of the network is much lower than the time duration of the changes in the forwarding layer, and the control plane will be aware of the announced state immediately\footnote{The control packets sent from the data plane to the control plane are very small in size comparing to the data packets. For example in case of transferring video files, no state changes happen in the data plane unless a video file is downloaded.  Since the size of the video files are much larger than the update packets (hundreds of megabytes comparing to a few bytes), the download time and thus the duration of state changes is much lower than the delay of the network for update packets. According to \cite{Gritter2001Architecture} the request round-trip latency in Akamai and Cisco are in the order of a few 10ms, while the download time for a 1GB movie using a very high speed internet of 100Mbps would take around 10s.
This is a practical assumption as well: if the delay in the network is longer than the duration of the changes, then a message sent to update the controller would be carrying obsolete information by the time it reaches the controller. Most practical systems are such that the time to notify the control is sufficient for the controller to use the information when it receives it. However, in any system, there is a chance that the actual state changes while the notification of the previous state is still underway, and there is always some distortion in the state representation at the control vs. the actual state in the forwarding plane.} (the alternative - that the state of the system changes as fast or faster than the control plane can be notified of these changes - is obviously unmanageable). Thus, the above errors may occur just when the forwarding plane does not send an update about a change. 

The main result now can be stated as a Lemma (with the proof in Appendix).

\begin{lemma}\label{lem:1}

If the ups and downs in the state of $X$ follow some distributions with means $\tau_X$ and $\theta_X$, respectively, then the minimum update rate $R_X(\epsilon_1,\epsilon_2)$ (number of update packets per second) satisfying the mentioned distortion criterion is given by
\begin{eqnarray}
&R_X(\epsilon_1,\epsilon_2) \geq
\frac{1}{\tau_X+\theta_X}  (2-\frac{\epsilon_1\frac{\theta_X}{\tau_X}}{\frac{\theta_X}{\tau_X+\theta_X}-\epsilon_2}-\frac{\epsilon_2\frac{\tau_X}{\theta_X}}{\frac{\tau_X}{\tau_X+\theta_X}-\epsilon_1})& \label{eq:rX} 
\end{eqnarray}
if $\frac{\epsilon_2}{1-\epsilon_2} < \frac{\theta_X}{\tau_X} < \frac{1-\epsilon_1}{\epsilon_1}$ and $\epsilon_2 \tau_X+\epsilon_1 \theta_X < \frac{\tau_X \theta_X}{\tau_X+\theta_X}$. Otherwise the distortion criteria is satisfied with no update at all.

\end{lemma}

Lemma \ref{lem:1} shows the minimum update rate for state of a single random variable X in the underlying plane so that an accepted amount of distortion is satisfied. The total rate and consequently the total protocol overhead for keeping the control layer informed of the forwarding layer is the combination of all the overheads needed for all the random processes of the underlying layer, which may be independent of each other or have some mutual impacts.

In the following section, we will use our model to formulate the control traffic needed in the interaction between caches and controller inside a sub-network.

The notations used here can be found in Tables \ref{tab:netmodel}-\ref{tab:scenario1}.

\section{Content Location in Cache Networks}
\label{sec:scenarios}

Information-centric networks \cite{Xylomenos2014Survey} usually employ routing-based \cite{Jacobson2009Networking} or resolution-based \cite{Dannewitz2013Network, DAmbrosio2011MDHT, Trossen2011PURSUIT, Koponen2007Dataoriented} methods for content discovery purposes.  In the routing-based discovery methods, like CCN, the required items are found exploring some areas of the network opportunistically or using other solutions like flooding. \cite{Wang2015ProDiluvian,Lee2015Content, Anastasiades2012Content} have studied these methods and proposed solutions to have the best performance. Resolution-based methods, on the other hand, require the control layer to know at least one location for each piece of data. PSIRP, DONA, and NetInf (partly) are some models which use the resolution-based methods. For instance, \cite{Trossen2011PURSUIT} attempts to set up a route to a nearby copy by requesting the content from a pub/sub mechanism. The pub/sub rendezvous point needs to know the location of the content. This is highly dynamic, as content can be cached, or expunged from the cache at any time. NDN \cite{Zhang2010Named} also assumes that the routing plane is aware of multiple locations for a piece of content\footnote{The routing (in NDN in particular) could know only one route to the content publisher or to an origin server and find cached copies opportunistically on the path to this server. But Fayazbakhsh et. al. \cite{Fayazbakhsh2013Less} have demonstrated that the performance of such an ICN architecture would bring little benefit over that of strict edge caching.}.

In a cache network, the addition/removal of an item (pieces of data which are requested and used by the users) to/from a cache may affect the timings of the other items in that cache; caching a piece of content somewhere may force another content out of the cache, and the caching policy will thus influence the network state (the existing items information), so we need to consider this effect in our calculations. It is worth noting that this framework may be used for CDNs as well, since the basics are the same, the point is that the update traffic for reporting the state of the caches in CDN would be very close to zero, since there are not a large number of changes in their states, unless the acceptable distortion is very low. We assume from now on that the Least-Recently-Used (LRU) replacement policy is used in the caches, as it is a common policy and has been suggested in some ICN architectures~\cite{Jacobson2009Networking}. However, based on \cite{Martina2014Unified}, other policies can be handled in a similar manner by generalizing the decoupling technique of Che's approximation \cite{Che2002Hierarchical}.

The request process also impacts the cache state, and we make the usual assumption that the items are requested according to a Zipf distribution with parameter $\alpha$; meaning that the popularity of an item $i$ is $\alpha_i=\frac{i^{-\alpha}}{\sum_{k=1}^M k^{-\alpha}}$, where $M$ is the size of the content set.

In the following sections we first introduce our framework to model the protocol overhead in section \ref{sec:framework}. Then, in section \ref{subsec:scenario1}, we use our model to study the total data retrieval cost including the control overhead and data downloading costs in a network of caches, where the nodes update the control plane of a domain (say, an AS) so as to route content to a copy of the cache within this domain if it is available. We denote the control plane function which locates the content for each request as the Content Resolution System (CRS).  

\subsection{Cache-Controller Interaction}
\label{subsec:scenario1}

Assume that we partition the network into smaller sub-networks each with its own control plane, such that all the nodes in each one of them have similar request patterns. A possible example of such partitioning are the Autonomous Systems (ASs) in the Internet.

Whenever a client has a request for an item, it needs to discover a location of that item, preferably within the AS, and it downloads it from there. To do so, it asks a (logically) centralized \emph{Content Resolution System (CRS)} by sending a Content Resolution Request (CRReq) or locates the content by any other non-centralized locating protocol. The Content Resolution Reply (CRRep) sent back to the client contains the location of the item, then the client starts downloading from the cache identified in CRRep.

If the network domain is equipped with a CRS, it is supposed to have the knowledge of all the caches, meaning that each cache sends its item states (local presence or absence of each item) to the CRS whenever some state changes.

Depending on the caching policy, whenever a piece of content is being downloaded, either no cache, all the intermediate caches on the path, or just the closest cache to the requester on the download path stores it in its content store, independently of the content state in the other caches, or refresh it if it already contains it.

We consider an autonomous network containing $N$ nodes (terminals), each sending requests for items $i=1,...,M$ with sizes $B_i$ according to a Poisson distributed process with rate of $\gamma_i$. The total request rate for all the items from each node is denoted by $\gamma=\sum_{i=1}^{M}\gamma_i$. Note that all the nodes in an AS have the same request pattern, i.e. content locality is assumed uniform in each AS\footnote{This assumption is widely used in works using the mathematical modeling for the networks \cite{Breslau1999Web}. This comes from the fact that 1) The requests coming from a specific region are very likely to follow similar patterns, because the users' interest in one area are highly correlated and can be predicted by having the information about just part of them \cite{bacstuug2015big}. For example, some certain news title  might be of special interest in a certain area, or some new TV series might be very popular in a certain country. 2) each user in this paper can actually be a hot spot or a base station, so a request generated from a node is not coming from one specific user but a group of users. So since we assume random users per station, then the assumption of uniform user locality is the best fitted assumption.}$^{,}$, and that the total request rate of each terminal is a fixed rate independent of the total number of nodes and items while the total requests for all the nodes is a function of $N$ (namely $N\gamma$). {If different users have different request distributions, then less cached contents will be reused, and thus there will be more changes in the cache states, and consequently more update traffic will be needed. The uniform content locality will give us the minimum required update rate.}

Suppose that there are $N_c$ caches in the system ($\mathcal{V}_c=\{v_1,...,v_{N_c}\}$) each with size $L_c$ that can keep (and serve) any item $i$ for some limited amount of time $\tau_i$, which depends on the cache replacement policy. Based on the the rate at which each item $i$ enters a cache and the time it stays there, each cache may have item $i$ with some probability $\rho_i$.  For simplicity, we assume that the probability distribution of the contents in all the caches are similar to each other. {We can easily extend to the case of heterogeneous caches at the cost of notation complexity. For instance, Theorem \ref{theo:1} below can be stated as a sum over all $N_c$ possible types of caches with $N_c$ different $\rho_i$s for each type of cache, instead of a product by $N_c$ of identical terms. Our purpose is to describe the homogeneous case, and let the reader adapt the heterogeneous case to suit her/his specific needs.}

In the following Theorem we want to compute the update rate for this system. Let $\bar{N_c}$ denote the number of caches where each downloaded piece of content is stored in (and thus need to send an update), either on the downloaded path, or off-path (Caching policy and network topology are the two factors that determine this number.). Thus, the rate of requests for item $i$ received by each cache is $\lambda_i=\gamma_iN \bar{N_c}/N_c$. 

\begin{table}[http]
	\begin{center}
		\begin{tabular}{|l|l|}
		  \hline
		   \small{Parameter} & \small{Definition} \\
		  \hline
		  \hline
		  $N$ & \small{Number of users} \\
		  $M$ & \small{Total number of items} \\
		  $N_c$ & \small{Number of content stores/caches} \\
		  $\bar{N_c}$ & \small{Number of caches storing the downloaded content} \\
		  $L_c$ & \small{Storage size per content store (bits)} \\
		  $B_i$ & \small{Average size of item $i$ (bits)} \\
			$\alpha_i$ & \small{Popularity of item $i$ (Zipf)}  \\
			$\gamma_i$ & \small{Total request rate for item $i$ per user}  \\
			$\gamma$  & \small{Total request rate per user} \\
			$\lambda_i$ & \small{Total request rate for item $i$ received per cache}  \\
			$\lambda$  & \small{Total request rate received per cache} \\
			\hline
			\end{tabular}
	\end{center}
	\caption{\textit{Parameters of the network model}}
	\label{tab:netmodel}
\end{table}

\begin{theorem}\label{theo:1}
The minimum total update rate for each item $i$ in the worst case is
\begin{eqnarray}
   &&R_{i}(\epsilon_1,\epsilon_2)\geq N_c\lambda_i (1-\rho_i) \nonumber \\
   &&\{2-\frac{\epsilon_1(1-\rho_i)}{\rho_i(1-\rho_i-\epsilon_2)}-\frac{\epsilon_2 \rho_i}{(1-\rho_i)(\rho_i-\epsilon_1)}\} \label{eq:ri}
\end{eqnarray}

if $\epsilon_1 < \rho_i < 1-\epsilon_2$ and $\epsilon_1 (1-\rho_i)+\epsilon_2 \rho_i < \rho_i(1-\rho_i)$. Otherwise no update is needed.
\end{theorem}

\begin{IEEEproof}
Let the random process $X$ in the forwarding plane denote the existence of item $i$ in a cache $v_j$ at time $t$, which is needed to be reported to the control plane (CRS). Let $\tau_{ij}$ denote the mean duration of time item $i$ spends in any cache $v_j$, and $\theta_{ij}$ denote the mean duration of time item $i$ not being in the cache $v_j$. 

In order to keep the CRS updated about the content states in the network, all the nodes have to send update packets regarding their changed items to CRS. All the assumptions of section \ref{sec:framework} are valid here. Thus, by replacing $\tau_X$ and $\theta_X$ in equation (\ref{eq:rX}) with $\tau_{ij}$ and $\theta_{ij}$ respectively, the result ($R_{ij}=R_X$) shows the minimum rate at which each cache $v_j$ has to send information about item $i$ to the CRS.

It can be seen that at the steady-state, the probability that cache $v_j$ contains item $i$ will be $\rho_{ij}=\frac{\tau_{ij}}{\theta_{ij}+\tau_{ij}}$. On the other hand, the total rate of generating (or refreshing) copies of item $i$ at each cache $v_j$, denoted by $\lambda_{ij}$, equals to $\frac{1}{\theta_{ij}}$.
Replacing the values of $\frac{\tau_{ij}}{\tau_{ij}+\theta_{ij}}$ and $\frac{1}{\theta_{ij}}$ in $R_{ij}$ with $\rho_{ij}$ and $\lambda_{ij}$ respectively, we obtain
\begin{eqnarray}
	&&R_{ij}(\epsilon_1,\epsilon_2) \geq \lambda_{ij}(1-\rho_{ij}) \nonumber \\
	&&\{2-\frac{\epsilon_1(1-\rho_{ij})}{\rho_{ij}(1-\rho_{ij}-\epsilon_2)}-\frac{\epsilon_2\rho_{ij}}{(1-\rho_{ij})(\rho_{ij}-\epsilon_1)}\}
\end{eqnarray}
	
for $\epsilon_1 < \rho_{ij} < 1- \epsilon_2$ and $\epsilon_2\rho_{ij}+\epsilon_1(1-\rho_{ij})<\rho_{ij}(1-\rho_{ij})$.

It is worth noting that we are not assuming any specific topology or caching policy here; the items may be cached on-path or off-path; just one cache may keep the downloaded content or a few caches may keep it. We are looking for the minimum amount of update packets in the worst case, which happens when each cache stores items independent of the items in other caches. It is obvious that topologies like a line of caches which result in strongly dependent caches are not in the scope of this paper. Thus, the total update rate for item $i$, is the sum of the update rates in all caches which is $R_i(\epsilon_1,\epsilon_2)=\sum_{j=1}^{N_c}R_{ij}(\epsilon_1,\epsilon_2)$. Recalling the assumption of (probabilistically) similar caches, we can drop the index $j$ and express the total update rate of item $i$ in terms of the probability of this item being in a cache.
This yields the result of equation~(\ref{eq:ri}) and the total update rate for all the items is the summation of these rates.
\end{IEEEproof}

\begin{table}[http]
	\begin{center}
		\begin{tabular}{|l|l|}
		  \hline
		  \small{Parameter} & \small{Definition} \\
		  \hline
		  \hline
			$\tau_i$  & \small{Average time item $i$ stays in a cache} \\
			$\theta_i$  & \small{Average time a cache does not have item $i$} \\
			$\rho_i$  & \small{Probability of item $i$ being in a cache} \\
			$\epsilon_{1,2}$ & \small{Distortion thresholds} \\
			$R_{ij}(\epsilon_1,\epsilon_2)$ & \small{Minimum rate at which each cache $v_j$} \\ & \small{must send update state of item $i$ to CRS so} \\ & \small{the defined distortion criteria is satisfied} \\
			$R_i(\epsilon_1,\epsilon_2)$ & \small{Minimum total update rate for item $i$ that} \\ & \small{satisfies the defined distortion criteria} \\
			$R(\epsilon_1,\epsilon_2)$ & \small{Minimum total update rate that} \\ & \small{satisfies the defined distortion criteria} \\
			\hline
	 \end{tabular}
	\end{center}
	\caption{\textit{Parameters used in cache-controller interaction}}
	\label{tab:scenario1}
\end{table}

\subsection{Model Evaluation and Simulation Results}
\label{subsec:eval}

\begin{figure}[http]
    \center
      \includegraphics[scale=0.22,angle=0]{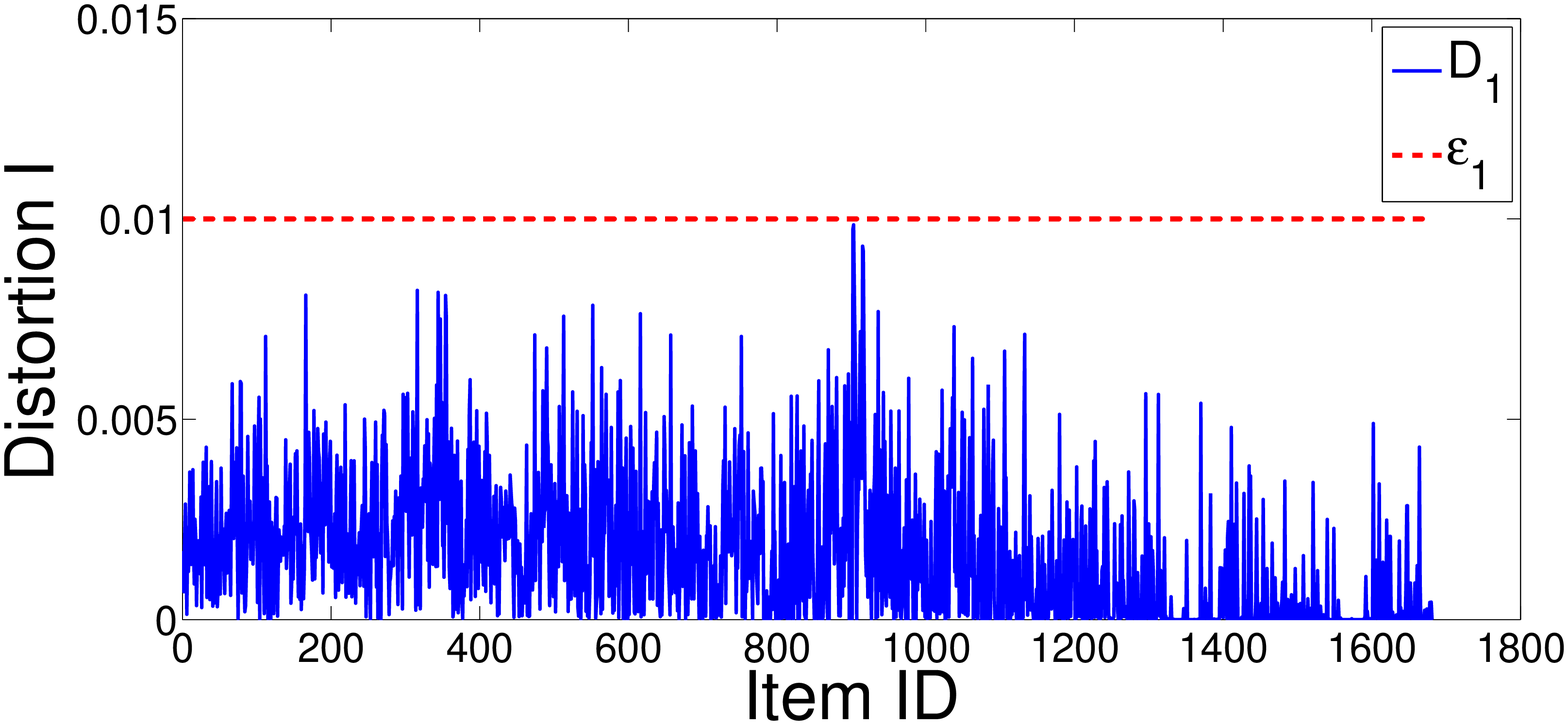}
      \includegraphics[scale=0.22,angle=0]{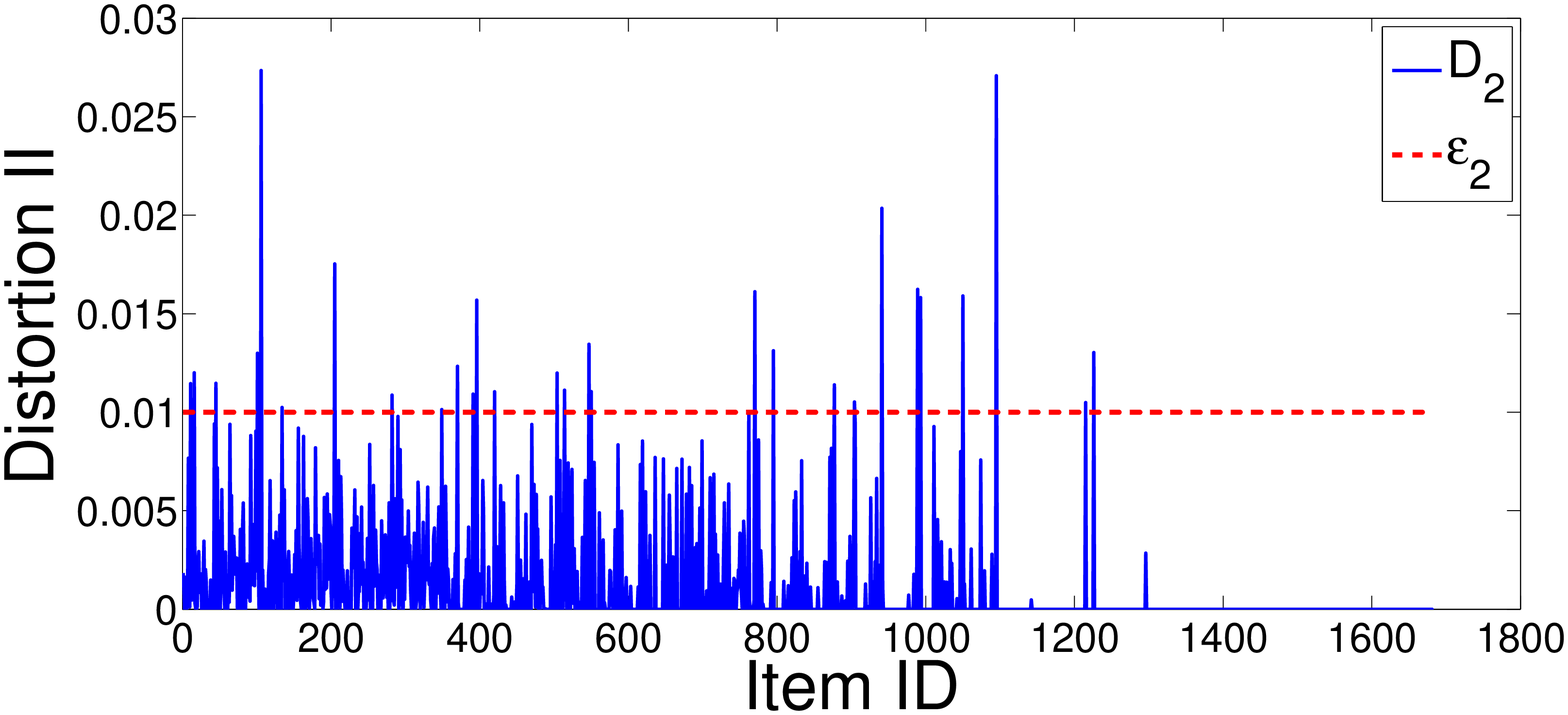}\\
        \includegraphics[scale=0.22,angle=0]{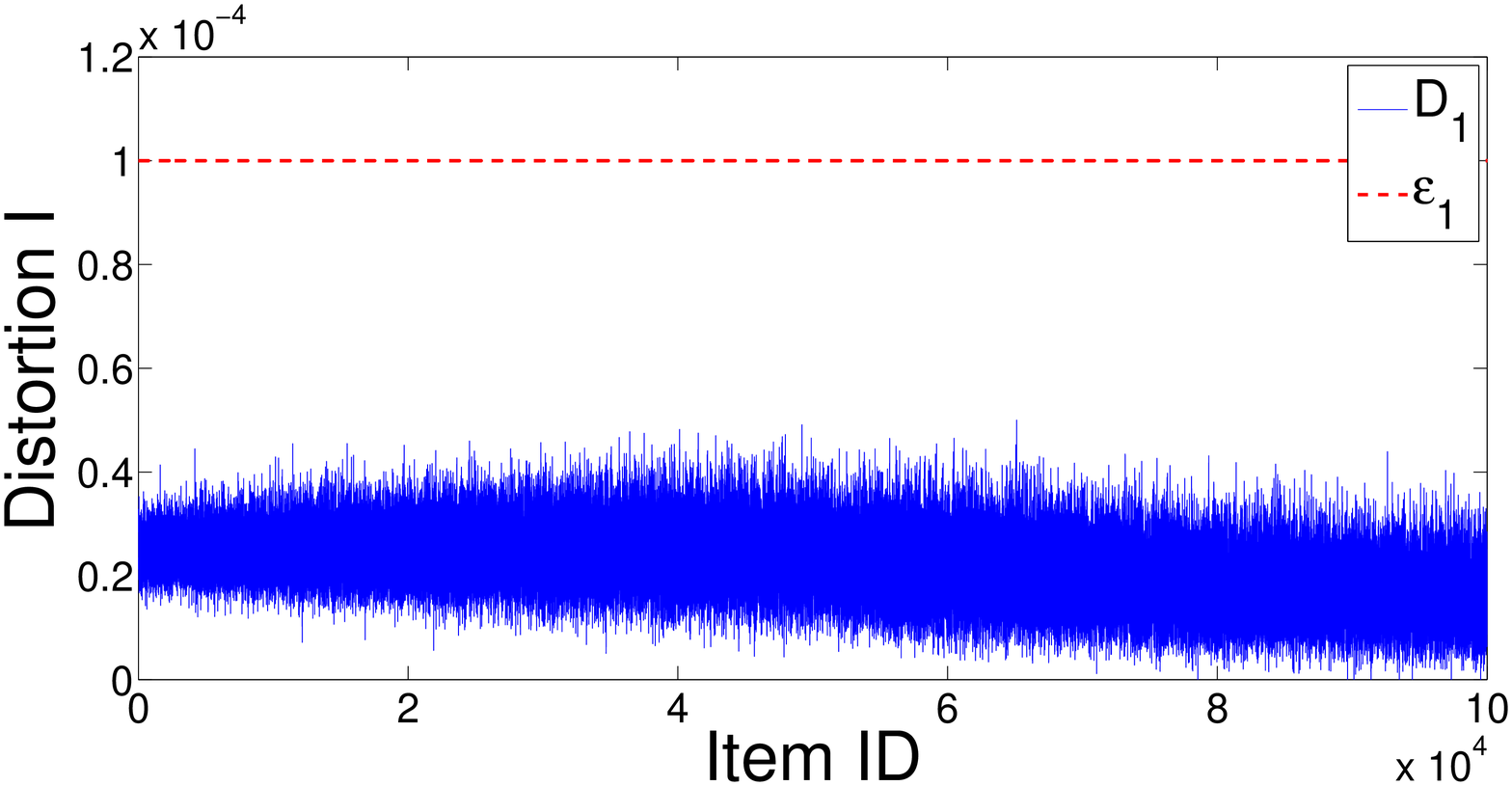}
      \includegraphics[scale=0.22,angle=0]{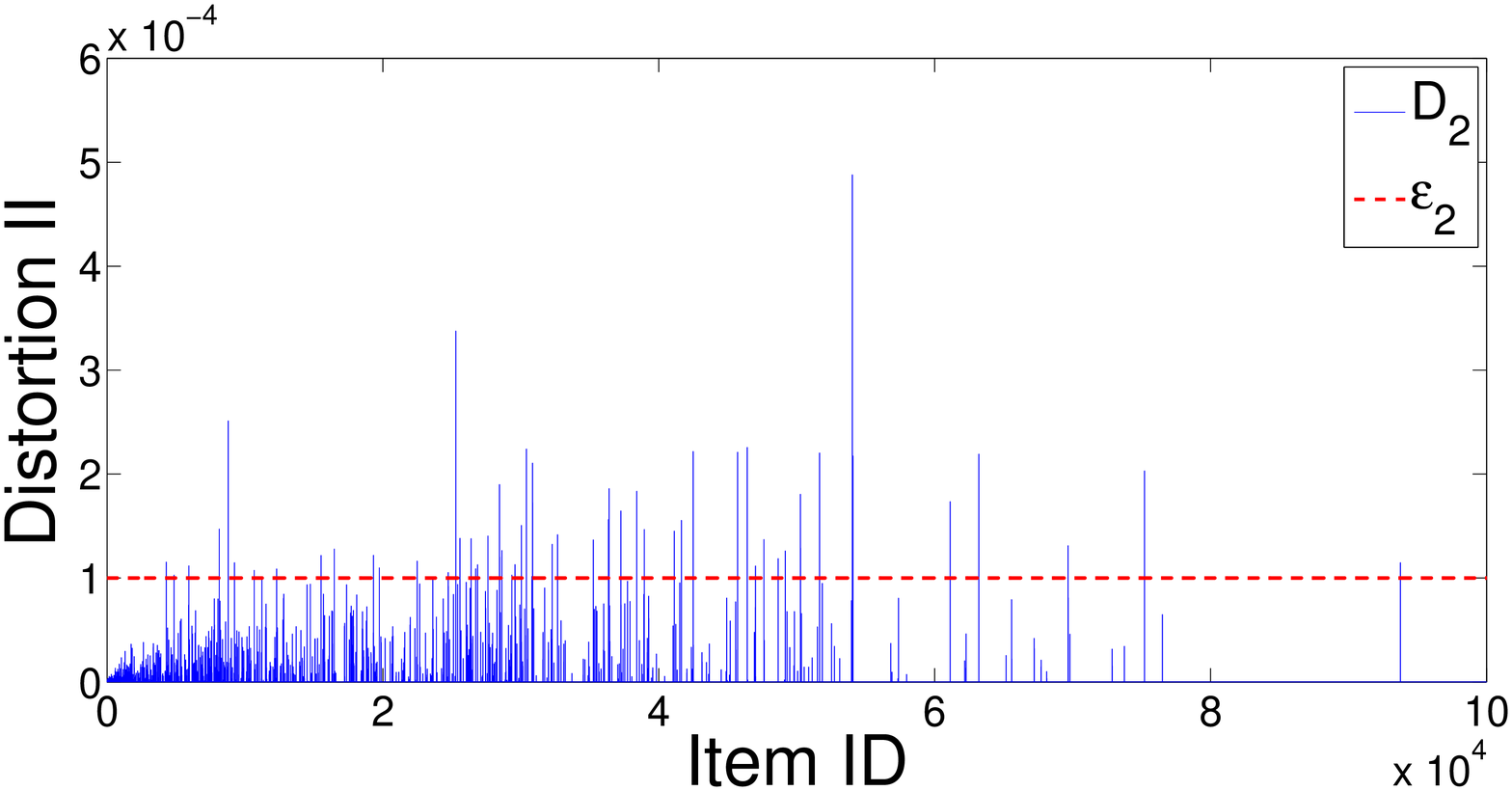}\\
      \caption{\textit{Measured distortion type I ($D_1$) and II ($D_2$) top) for MovieLens dataset ($100K$ requests for $M=1,682$ movies), with $\epsilon_1=\epsilon_2=0.01$ as accepted distortions, and bottom) for synthetic dataset ($10M$ Poisson requests for $M=100K$ Zipf distributed movies with skew factor $\alpha=0.7$), with $\epsilon_1=\epsilon_2=10^{-4}$ as accepted distortions.}}
    \label{fig:EvalDist1}
\end{figure}

To figure out how the calculated rates perform in practice and evaluate our model, we simulate an LRU cache with capacity $L_c=20$ items. We use the MovieLens dataset \cite{Harper2015MovieLens}, which contains $100,000$ ratings together with their time stamps collected for $M=1,682$ movies from $943$ users during a seven-month period. We took the ratings as a proxy for content requests, assuming that the users who reviewed the movie have requested them shortly prior to the review. In these simulations we first estimate the item availability in the cache $\rho_i$ (by dividing the total time that item is in the cache by the total simulation time), then using the estimated $\rho_i$ and according to equations (\ref{eq:U1}) and (\ref{eq:U2}), we calculate the update rate in case of a change. Then we run the simulation for $100,000$ requests. In this simulation we update the CRS according to the calculated rates, which can be interpreted as the chances of update, whenever a change occurs in the cache. Then we measure the total time that the CRS information does not match the actual cache state for each item, and calculate the average generated distortion during 10 rounds of simulation. The top figures in Figure \ref{fig:EvalDist1} illustrate the results for the case where $\epsilon_1=\epsilon_2=0.01$.

Since, according to \cite{Koponen2007Dataoriented} and \cite{bankoski2017ICNC}, the number of data objects is very large, and is becoming larger, we repeated similar evaluation with a relatively large synthetic dataset, containing $10$ million Poisson requests for contents picked from a catalog of $100K$ movies, according to a Zipf distribution with skew parameter $\alpha=0.7$. Bottom figures in Figure \ref{fig:EvalDist1} show the results for the synthetic dataset allowing $\epsilon_1=\epsilon_2=10^{-4}$ distortion accepted (larger number of contents leads to lower cache availability, thus we allowed lower distortion here).

It must be noted here that we are estimating $\rho_i$ based on the past cache states, so it is not the exact $\rho_i$.Thus the generated distortion may exceed the tolerable values for some items, while they are in the safe zone for the others. It is observed that for a large portion of the items the distortion type I satisfies the distortion criteria. Distortion type II, however, does not satisfy the distortion criteria for more items. The reason is that the calculated update rates are strongly dependent on the availability of the items in the cache and any small error in the estimation of $\rho_i$ may lead to some extra distortions. Since the $\rho_i$'s are mostly very small, not updating just one type II change may cause an error which remain in the system for a long time, and thus creating a large distortion.

\begin{figure}[http]
    \center
      \includegraphics[scale=0.24,angle=0]{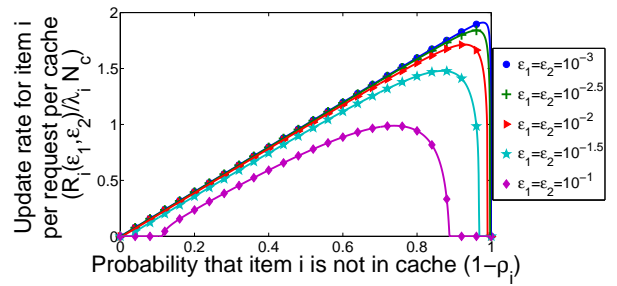}\\
      \caption{\textit{Number of needed updates for item $i$ sent from all caches to CRS per generated request for that item versus $1-\rho_i$ for different distortion criteria.}}
    \label{fig:updatemiss}
\end{figure}

Figure \ref{fig:updatemiss} illustrates the number of needed updates per generated request for each item $i$ in the network $\frac{R_i}{N_c\lambda_i}=\frac{R_i}{N\gamma_i}$ when the caches does not contain it with a known probability ($1-\rho_i$). The only variable parameters in this graph are $\epsilon_1$ and $\epsilon_2$. The higher distortion we tolerate, the less update announcements for each item $i$ we need to handle. Moreover, the number of items which need some updates is decreasing when higher distortions are accepted. As can be seen the update rate starts from zero for those items which are in the cache with high probability. Status of these items are permanently set to $'1'$ in CRS, and no update is needed. At the other end of the graph, for the items which are almost surely not in the cache, The presence probability is close to zero ($\rho_i=0$ and thus $1-\rho_i=1$)), and the status of those items can be permanently set to $'0'$ in the control plane, thus the caches don't need to send any more information regarding those items to the control plane. Therefore, again no update is needed.

\begin{figure}[http]
    \center
      \includegraphics[scale=0.24,angle=0]{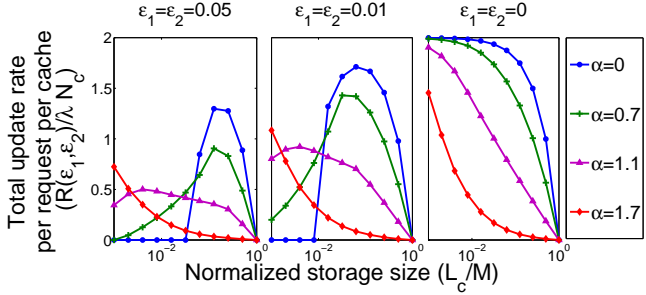}\\
      \caption{\textit{Total cache-CRS update rate (Updates per generated request per cache) for different cache storage capacities and different acceptable distortions. Content set contains $M=1,000$ contents.}}
    \label{fig:lruLc}
\end{figure}

The probability $\rho_i$ is strongly dependent on the cache replacement policy. We consider LRU as the cache replacement method used in the network. Clearly, in LRU caches (similarly in other policies like FIFO, LFU, etc.) $\rho_i$ is just a function of the probability of item $i$ coming to the cache ($\alpha_i$), and the cache capacity ($L_c$).
Figure \ref{fig:lruLc} shows the changes of the total update rate (scaled by $\frac{1}{\lambda N_c}$) versus the cache storage size, in a network of LRU caches, such that the distortion criteria defined by $(\epsilon_1,\epsilon_2)$ is satisfied. In this simulation $M=10^3$. Note that each change in a cache consists of one item entering into and one other item being expunged from the cache, therefore if no distortion is tolerable, this rate will be $2$ updates per change per cache. 

It can be observed that for very small storage sizes and small popularity index, almost each incoming item changes the status of the cache and triggers an update. When the storage size is still very small, the caches do not provide enough space for storing the items and reusing them when needed, so increasing the size will increase the update rate. At some point, the items will move down and up in the cache before going out, so increasing the storage size more than that will
reduce the need to update. However, if the popularity index is large, then increasing cache size from the very small sizes will decrease the need to update since there are just a few most popular items which are being requested.

Moreover, as it is expected, when more distortion is tolerable, the CRS needs fewer change notifications. However, if the cache size is too big, or the popularity exponent is too high, fewer changes will occur, but almost all the changes are needed to be announced to the CRS. On the other hand, for small cache sizes accepting a little distortion will significantly decrease the update rate.

\section{Application to Cost Analysis}
\label{sec:costanalysis}

In this section we use Theorem~\ref{theo:1} to study trade-offs involved in updating the content control layer. More specifically, we try to calculate the bounds on the total cost (required bandwidth for download + CRS update) and look at the trade-offs between the cost, the size of the information chunks, the number of caches, and the size of caches.

\subsection{Network Model}

\begin{figure}[http]
    \center
      \includegraphics[scale=0.44,angle=0]{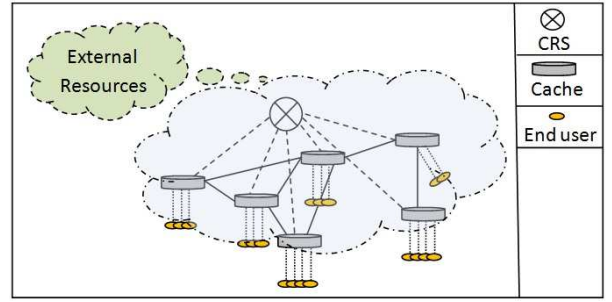}\\
      \caption{\textit{Network Model.}}
    \label{fig:netmodel}
\end{figure}

Figure \ref{fig:netmodel} illustrates the network model studied in this section. This model consists of entities in three substrates: users are located in first layer; a network of caches with the CRS on the second level; external resources (caches in other networks, Internet, etc.) on the third.

We need to define the relative costs of the different actions. We assume that the state update process for item $i$ has a per bit cost of $\xi^{up}_i$ for sending data from the cache to CRS, and per bit cost of $\xi^{extup}_i>\xi^{up}_i$ for sending data from one CRS to another one. On the other hand, the requested piece of content $i$ may be downloaded from a cache inside the network with some per bit cost $\xi^{int}_i$, or it may be downloaded from an external server with some other cost $\xi^{ext}_i>\xi^{int}_i$. These costs may be a function of the number of hops in the network. Note that the exact costs for each cache are determined based on the network topology, and may not be the same for all the caches. In this paper we use the average cost over all the caches \footnote{The other option for defining the distortion and correspondingly cost is the worst case, which will map to the maximum update cost. 
There are many few caches that may undergo some maximum number of changes and need some maximum update transfers, thus, this is clearly not illustrating the performance of a cache network correctly. We have decided to work with the average method, which is very common in the literature (References \cite{Wang2008Link,Wang2012Cost,Hong2009Impact}) and we believe it can represent the performance of the entire network better in our specific application and many others.}.

\subsection{Total Cost in Cache-Controller Interaction}

\begin{lemma}\label{lem:2}
The length of each update packet for  content $i$ is
\begin{equation}
   l_i \geq \log N_c-\log \frac{\lambda_i(1-\rho_i)}{\sum_{i=1}^M\lambda_k(1-\rho_k)} +1
\end{equation}
\end{lemma}

\begin{IEEEproof}
Each update packet contains the ID of the cache issuing the query, the ID of the updated item and its new state. There are $N_c$ caches in the network, hence, $\log N_c$ bits are needed to represent the cache. Item $i$ is updated with probability $\beta_i=\frac{\lambda_i(1-\rho_i)}{\sum_{i=1}^M\lambda_k(1-\rho_k)}$, which results in a code length of at least $-\log \beta_i$ bits. Thus, the length of each update packet is $l_i\geq \log N_c-\log \beta_i +1$.
\end{IEEEproof}
 
\begin{lemma}\label{lem:3}
The total update cost in the defined network is 
\begin{eqnarray}
   \varphi^{up} = \sum_{i=1}^M R_i(\epsilon_1,\epsilon_2)  l_i  \xi^{up}_i.  \label{eq:l3}
\end{eqnarray}

where $R_i(\epsilon_1,\epsilon_2)$ is the minimum rate at which the update state of item $i$ must be reported to CRS so that a distortion criteria defined by $(\epsilon_1,\epsilon_2)$ is satisfied.
\end{lemma}

\begin{IEEEproof}
Each cache sends update packets at rate $R_i(\epsilon_1,\epsilon_2)$ to provide its CRS with the state of item $i$ in its local content store. Each update packet contains $l_i$ bits, and there is a per bit cost of $\xi^{up}_i$ for the update packets. Therefore, the cost for updating information about item $i$ in the sub-network is $\varphi^{up}_i = R_i(\epsilon_1,\epsilon_2)  l_i  \xi^{up}_i$, and the total update cost is $\varphi^{up} = \sum_{i=1}^M R_i(\epsilon_1,\epsilon_2)  l_i  \xi^{up}_i$.
\end{IEEEproof}

\begin{lemma}\label{lem:4}
The total download cost in the defined network is 
\begin{eqnarray}
   \varphi^{dl} = \sum_{i=1}^M N\gamma_i  B_i  ((P_i-\rho_i)  \xi^{int}_i + (1-P_i)  \xi^{ext}_i)
\end{eqnarray}

if $\epsilon_1 < \rho_i < 1-\epsilon_2$ and $\epsilon_1 (1-\rho_i)+\epsilon_2 \rho_i < \rho_i(1-\rho_i)$. Otherwise no update is needed.
\end{lemma}

\begin{IEEEproof}
The requested piece of content $i$ may be downloaded from the local cache with cost $0$ (with probability $\rho_i$ of being in this cache), from another cache inside the same network with some per bit cost $\xi^{int}_i$ (with a probability we denote by $P_i-\rho_i$, where $P_i$ is the probability that content $i$ is within the AS's domain), or it must be downloaded from an external server with some other cost $\xi^{ext}_i$ (with probability ($1-P_i$)). Obviously, $\rho_i \leq P_i \leq 1$. Thus, the download cost for item $i$ with size $B_i$ bits for each user in the sub-network is

\begin{equation}
\varphi^{dl}_{ij} = \gamma_i  B_i  ((P_i-\rho_i)  \xi^{int}_i + (1-P_i)  \xi^{ext}_i),
\end{equation}

The total download cost for item $i$ is $\varphi^{dl}_i=N\varphi^{dl}_{ij}$, and the total download cost for all the items is the summation of $\varphi^{dl}_i$'s over all the contents. 
\end{IEEEproof}

\begin{theorem}\label{thm:2}
The total cost in the defined network including update and download costs is 
\begin{eqnarray}
   \varphi &=& \sum_{i=1}^M N\gamma_i  B_i  ((P_i-\rho_i)  \xi^{int}_i + (1-P_i)  \xi^{ext}_i) \nonumber \\
   &+& \sum_{i=1}^M R_i(\epsilon_1,\epsilon_2)  l_i  \xi^{up}_i.
\end{eqnarray}
\end{theorem}

\begin{IEEEproof}
Adding Lemmas \ref{lem:3} and \ref{lem:4} proves the Theorem.
\end{IEEEproof}

It can be seen that the total cost is strongly dependent on where each query is served from, and consequently on the probability of each item being internally served ($P_i$). This probability depends on the probability of that item being in an internal cache, which is in turn a function of the caching criteria and the replacement policy.
Lemma \ref{lem:5} presents some bounds on $P_i$ based upon the allowed distortion. The proof can be found in appendix.

\begin{lemma}\label{lem:5}
The probability that each content $i$ is internally served is bounded by 
\begin{equation}
[1-(1-\rho_i+\epsilon_1)^{N_c}]^+\leq P_i \leq 1-(1-\rho_i)^{N_c}.
\end{equation}
where $[x]^+=max(x,0)$,
\end{lemma}

Note that the above $P_i$ may take any value in the calculated bounds depending on the value of $\rho_i$. For example if $\rho_i<\epsilon_1$ then $D_{1_i}=\rho_i$, and $P_i=0$, which is the lowest value of this bound. On the other hand, if $\rho_i>1-\epsilon_2$ then $D_{1_i}=0$, and $P_i=1-(1-\rho_i)^{N_c}$, which is the highest value in this bound. All the other values of $\rho_i$ will lead to $P_i$ between those two boundaries.

These two extreme cases of $P_i$ result in some bounds on the download cost. Let $\varphi^{dl}_L$ and $\varphi^{dl}_H$ denote the lower and upper bounds of the download cost corresponding to the upper and lower bounds of $P_i$, respectively, and  $\varphi_L$ and $\varphi_H$ denote the lower and upper bounds of the total cost. Note that for small values of the tolerable distortion $\epsilon_1$ the upper and lower limits of $P_i$ and correspondingly the bounds of download cost are very close to each other.

\begin{figure}[http]
    \center
      \includegraphics[scale=0.22,angle=0]{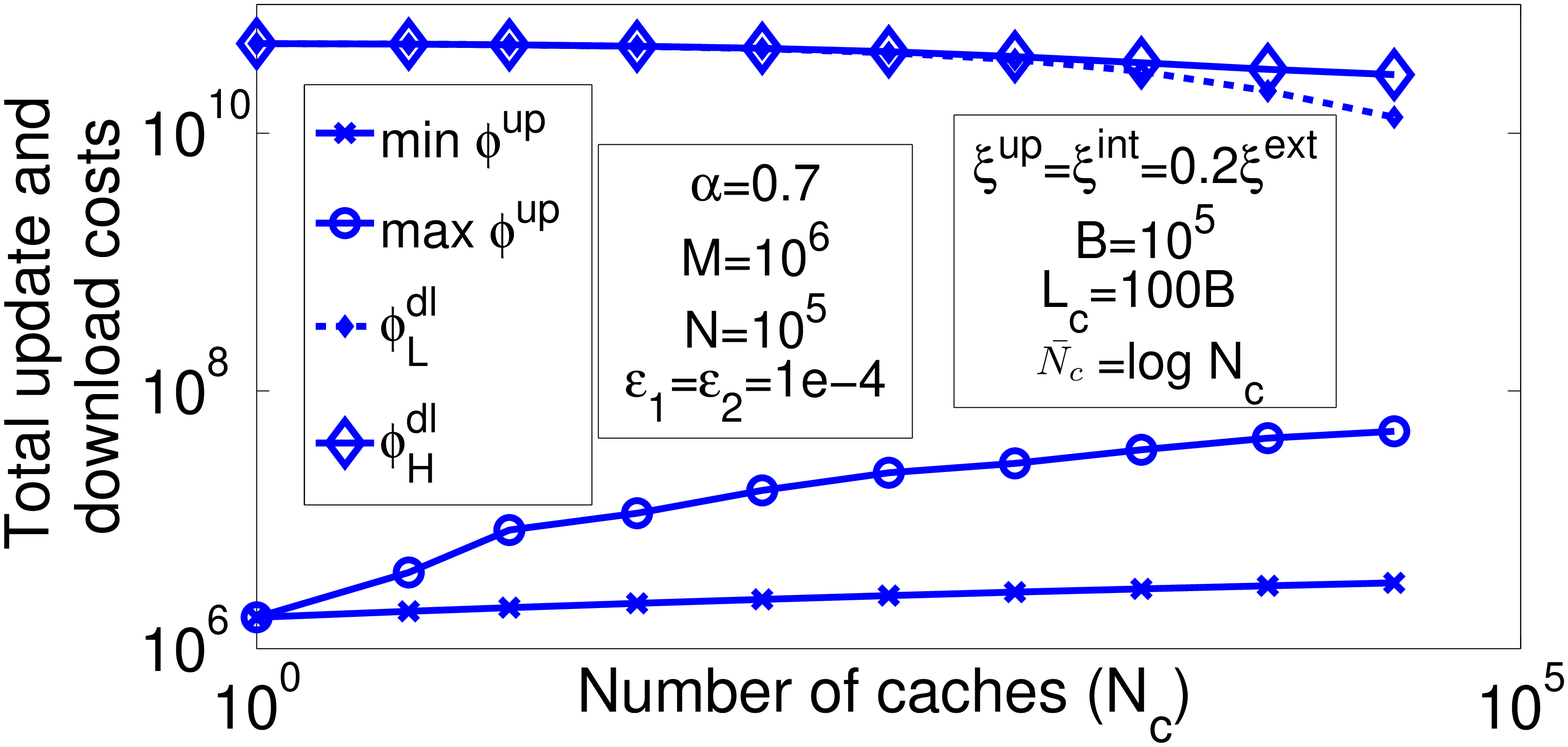} 
      \includegraphics[scale=0.22,angle=0]{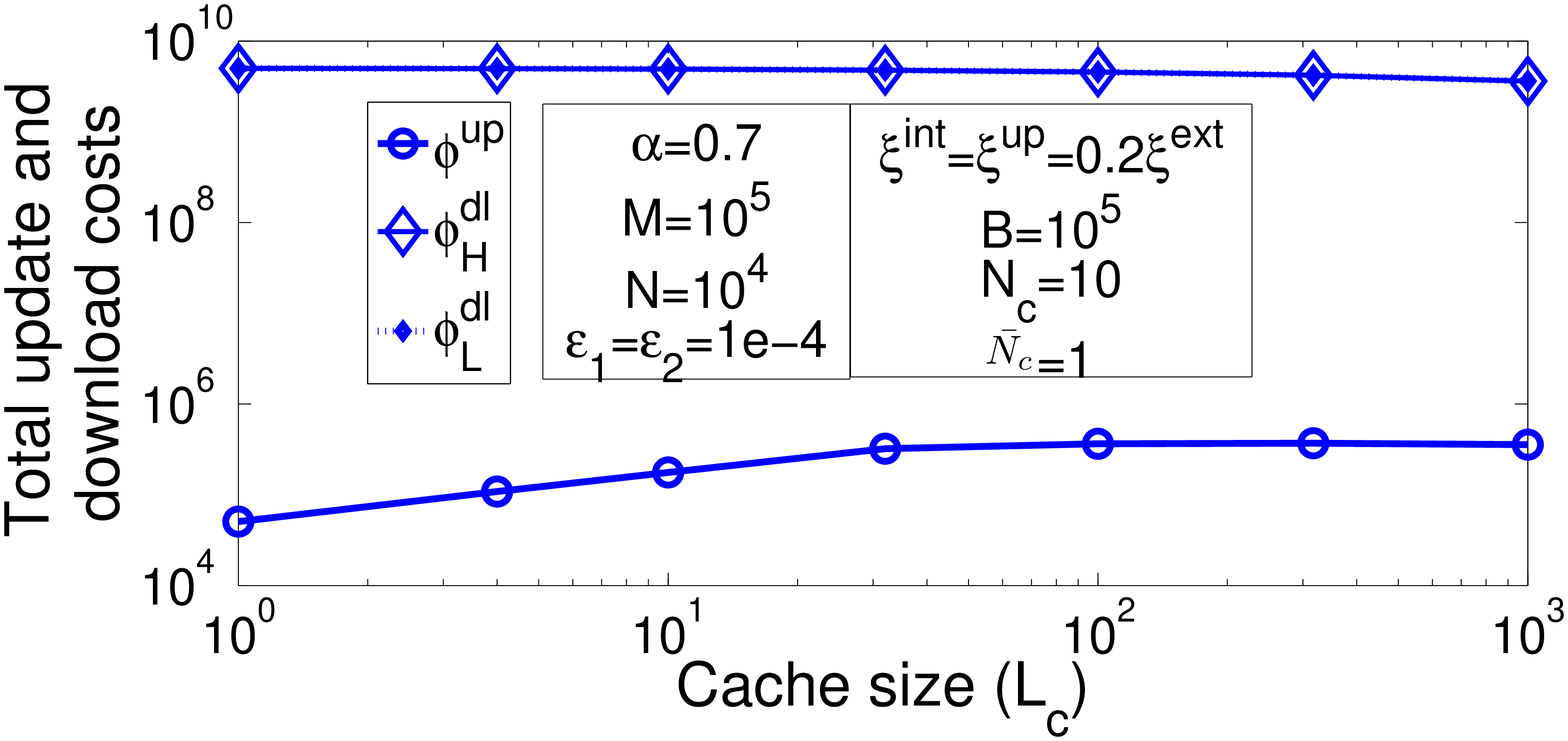} \\
      \caption{\textit{Total update cost (minimum and maximum of $\varphi^{up}$) and total download cost (lower and upper bounds, $\varphi^{dl}_L$ and $\varphi^{dl}_H$, left) vs. the number of caches ($N_c$), when each item is $B=10^5$ units long, the storage size per cache is fixed ($L_c=100$ items), and each downloaded item is stored in $\bar{N_c}=\log N_c$ caches, and right) vs. the cache size ($L_c$), when each item is $B=10^5$ units long, the number of caches is fixed ($N_c=10$), and each downloaded item is stored in $\bar{N_c}=1$ cache.}}
    \label{fig:updlcost1}
\end{figure}

Figure \ref{fig:updlcost1} the left plot illustrates the changes of update and download costs in a network with a content set of a total of $1$ million contents, when the size of each cache is limited to $L_c=100$ contents. The length of the data packets is assumed to be $100KB$ in average, while the update packets are $l_i$ bytes each. Note that increasing the data (or update) packet lengths will increase the download (or update) cost linearly.

Here we assume that whenever an item is downloaded, it is stored in $\bar{N_c}=\log N_c$ caches, which have to report the changes to the controller\footnote{This happens in largely used network models like binary tree or grid topology, when all the caches on the download path store the content.}. If these caches are selected randomly, the total update rate would be $\bar{N_c}$ times the rate of update of each cache resulting in max $\varphi^{up}$. On the other side, if they are completely dependent, for example if all the caches on the download path keep it, then just one update may be enough, resulting in min $\varphi^{up}$. So depending on the caching policy, the update cost will be something between min and max $\varphi^{up}$. 

The request rate received by each cache is inversely proportional to the number of caches (the request rate per user is assumed to be fixed and independent of $N_c$), and the update packet length increases logarithmically with the number of caches. The total update rate per cache is almost linearly decreasing with $N_c$, hence the minimum of the total update rate, or the total update rate if just one cache keeps the downloaded item, will almost be stable when $N_c$ varies (changes are in the order of $\log N_c$). The maximum total update rate will linearly change with the number of copies $\bar{N_c}$ per download. Thus, increasing the number of caches in the network increases the update cost by a factor of at least $\log N_c$ and at most $\bar{N_c}\log N_c$.

Increasing $N_c$, however, increases the probability of an item being served internally and thus decreases the download cost. 
Nevertheless, as it can be observed, the rate of decrease is so low that it can be assumed as stable.
 
In the right plot of figure \ref{fig:updlcost1}, we fix the number of caches in the AS ($N_C=10$) and study the effects of cache storage size on the update and total cost. Increasing the cache size, simply increases the probability of an item being served internally and decreases the download cost. Again similar to the left plot, the rate of changes in the download cost is very low. As expected, on the other side, the update cost shows an increase when increasing the storage size. Looking at each cache, very small cache size leads to very large durations where that item is not in that cache and consequently, the update rate would be low. Increasing the storage size will increase the probability of that item being in the cache, and thus increases the update rate. If we let more cache storage, this increase reaches its highest value for a certain value of cache size, and for larger values of cache beyond a threshold, the item is in the cache most of the time. Therefore, we need less updates and increasing the cache size will increase the duration of the item being in the cache leading to fewer update messages. Since the total cost mostly depends on the download cost, by increasing the cache size, this value reaches its minimum value.

\begin{figure}[http]
    \center
      \includegraphics[scale=0.24,angle=0]{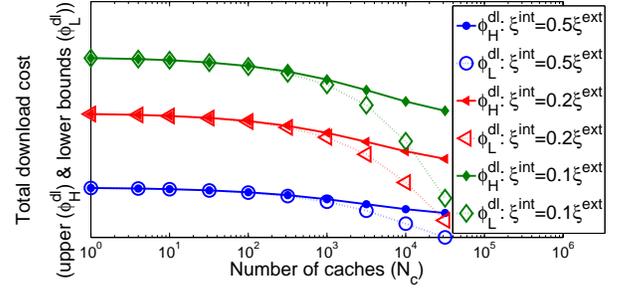} \\
      \caption{\textit{Total download cost (lower and upper bounds, $\varphi^{dl}_L$ and $\varphi^{dl}_H$) vs. the number of caches ($N_c$), for different download cost values, when each item is $B=10^5$ units long, and the storage size per cache is fixed ($L_c=100$ items).}}
    \label{fig:dlcost1}
\end{figure}

It is worth noting here that the cost of download from another AS has been assumed $5$ times bigger than the cost of download from inside the AS, which in turn is assumed to be the same as the update cost per bit. Figure \ref{fig:dlcost1} shows the impact of the external and internal costs on the total download cost. Higher external costs result in higher total download costs, as it is expected, and show more decrease rate when the number of caches increases. Thus, if the external download cost comparing to the internal download cost is very high, having more caches may make sense, although, the total cost decrease rate is still very low.
 
Another important result shown by figures \ref{fig:updlcost1} and \ref{fig:dlcost1} is that having big data packets, the download cost is always much higher than the update cost, which is reasonable. In other words, having the resolution-based content discovery when the data packets are large, will add just negligible cost to update the controller.  In the following we study the affect of chunk-based caching to obtain some insight on the reasonable size of chunks, such that the update cost remains negligible.

\begin{figure}[http]
    \center
      \includegraphics[scale=0.24,angle=0]{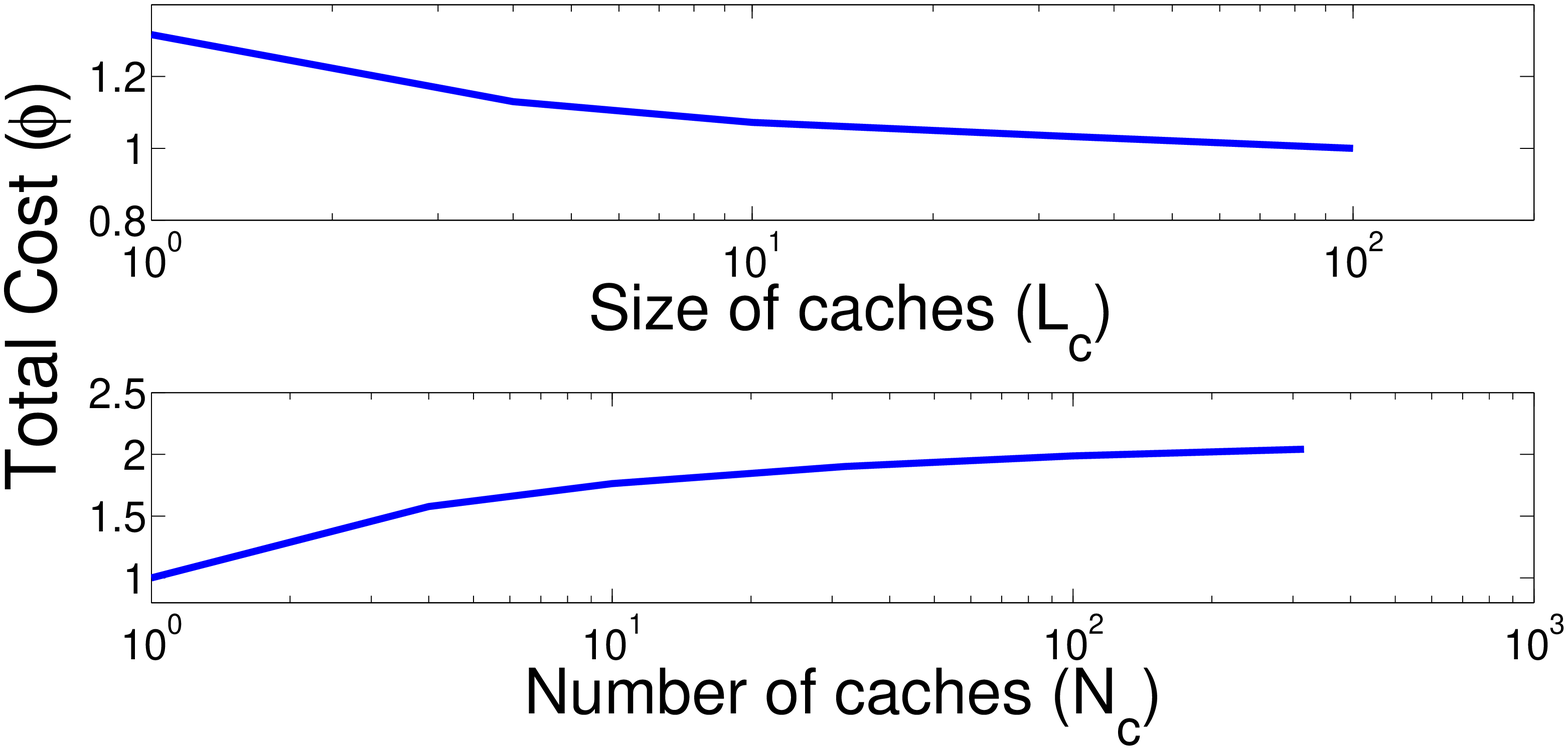}\\
      \caption{\textit{Total cost ($\varphi$),when the total storage size ($N_c L_c$) is fixed and equal to half of the catalog size, vs. top) the size of caches ($L_c$), and bottom) the number of caches ($N_c$).}}
    \label{fig:totcostfixstorage}
\end{figure}

The top plot  in figure \ref{fig:totcostfixstorage} shows the total cost versus the number of caches, when the LRU cache replacement policy is used and the total storage of the cache sublayer is limited to half of the total number of items. The parameters are set as follows: $M=10,000,\ \epsilon_{1,2}=1e-4,\ B_i=100K,\ \xi^{int}_i=1,\ \xi^{ext}_i=5,\ \xi^{up}_i=1$, and $\alpha=0.7$. Since the lower and upper bounds of the total cost are very close to each other, we just plot the upper limit here.
In the bottom plot of this figure, the total cost is plotted versus the size of each cache. It can be observed that with a fixed total storage size, concentrating all the caches in one node and increasing the size of it will lead to better overall performance (least cost). Note that in these figures the total cost value shown is just a relative value, and not the exact one.

\subsection{Optimized Cache Management}
\label{subsec:optimumi}

In previous section, the total cost in the described cache network was derived and the impacts of the number or size of the content stores on this cost was studied. We now turn our attention to minimizing the total cost for given $N_c$ and $L_c$.

Under a Zipf popularity distribution, many rare items will not be requested again while they are in the cache under the LRU policy. We can rewrite the total cost if the caches only keep the items with popularity from $1$ up to $i^*$.
\begin{eqnarray}
\varphi &=& \sum_{i=1}^{i^*} N\gamma_i  B_i  ((P_i-\rho_i)  \xi^{int}_i + (1-P_i)  \xi^{ext}_i) \nonumber \\ &+& \sum_{i=i^*+1}^M  N\gamma_i  B_i  \xi^{ext}_i + \sum_{i=1}^{i^*} R_i(\epsilon_1,\epsilon_2)  l_i  \xi^{up}_i  \label{eq:phiwthr}
\end{eqnarray}

Now just $i^*$ different pieces of content may be stored in each cache. This changes the probability of an item $i=1,...,i^*$ being in a cache ($\rho_i$), which in turn changes $P_i$ and $R_i$.

\begin{figure}[http]
    \center
      \includegraphics[scale=0.24,angle=0]{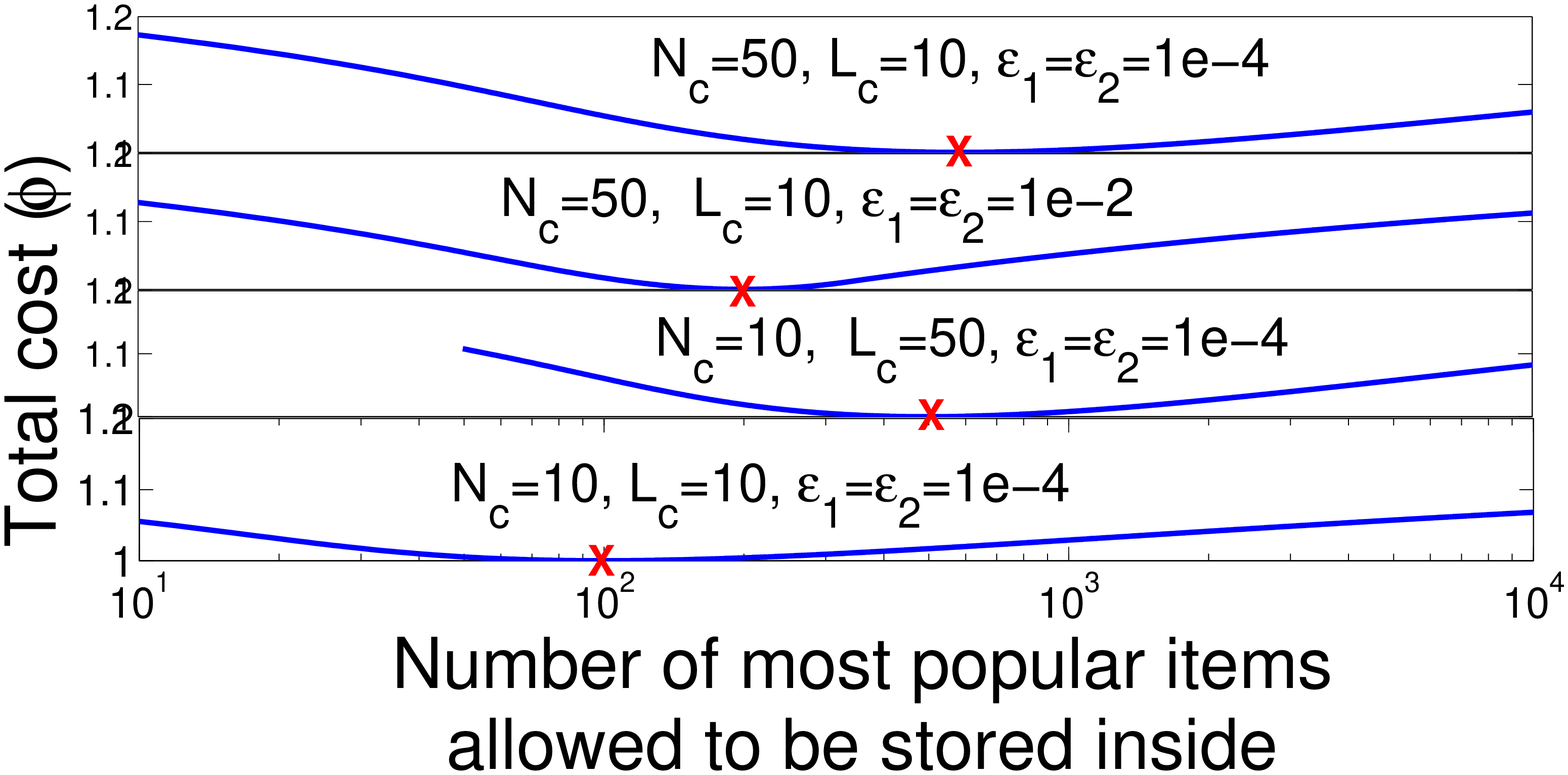}\\
      \caption{\textit{Total cost when just $i^*$ most popular items are allowed to be stored inside caches ($N=10^3,M=10^4,B=10^5,\alpha=0.7$).}}
    \label{fig:optIstar}
\end{figure}

Figures \ref{fig:optIstar} demonstrates the total cost versus the caching popularity threshold $i^*$, for different number and size of content stores, and acceptable distortions.

If just a very small number of items (small $i^*$) are kept inside cache layer, then the download cost for those which are not allowed to be inside caches will be the dominant factor in the total cost and will increase it. On the other hand, if a lot of popularity classes are allowed to be kept internally, then the update rate is increased and also the probability of the most popular items being served internally decreases, so the total cost will increase. There is some optimum caching popularity threshold where the total cost is minimized. This optimum threshold is a function of the number and size of the stores, distortion criteria, per bit cost of downloads and updates. 

The benefit of the optimized solution also varies depending on the mentioned parameters. For example according to figure \ref{fig:optIstar}, the optimized solution can have $17\%$ reduction in cost in case when $N_c=50,L_c=10,\epsilon_1=\epsilon_2=1e-4$, while this cost reduction is just $7\%$ when $N_c$ is five times smaller.

To find the optimal $i^*$, assume that all the items have the same size ($B_i=B$) and the per bit costs is fixed for all popularity classes ($\xi_i^{int}=\xi^{int}$, $\xi_i^{ext}=\xi^{ext}$, $\xi_i^{up}=\xi^{up}$). We can rearrange equation (\ref{eq:phiwthr}).
\begin{eqnarray}
\varphi=\varphi_1-\varphi_2+\varphi_3 \label{eq:tobemin}
\end{eqnarray}

where $\varphi_1=B N\gamma \xi^{ext}$ is the total cost if no cache exists and all the requests are served externally;
$\varphi_2=BN\gamma(\xi^{ext}-\xi^{int})\sum_{i=1}^{i^*} \alpha_i P_i+BN\gamma \xi^{int} \sum_{i=1}^{i^*} \alpha_i \rho_i$ corresponds to the benefit of caching (cost reduction due to caching); and $\varphi_3=\xi^{up} \sum_{i=1}^{i^*} R_i(\epsilon_1,\epsilon_2)l_i$ is the caching overhead cost due to the updates. The last term is the cost we pay because of caching (updates).  We need to calculate the value of $i^*$ such that the cost of caching is dominated by its advantage; i.e. we need to maximize $\varphi_2-\varphi_3$.

This can be done using numerical methods which will lead to a unique $i^*$ for each network setup (fixed parameters). However, the network characteristics and the request pattern are changing over time, so it seems that it is better to have a mechanism to dynamically optimize the cost by selecting the caching threshold ($i^*$) according to the varying network features.

In such a mechanism, the CRS can keep track of requests and have an estimation of their popularity. For those requests which are served locally the CRS can have an idea of the popularity based on the updates that receives from all the caches; i.e. the longer an item stays in a cache, the more popular it is. It can also take into account the local popularity of the items. The CRS can then dynamically search for the caching threshold which minimizes the total cost by solving equation (\ref{eq:tobemin}). Once the CRS determines which items to keep internally, it will set/reset a flag in each CRRep so that the local cache knows to store or not to store the requested piece of content.

\section{Conclusion and Future Work}
\label{sec:conclusion}

We formulated a distortion-based protocol overhead model. Some simple content distribution networks were then considered as examples to show how this framework can be used, and based on this model the overhead of keeping the control plane informed about the states of the contents in these networks was calculated. It was confirmed that with big data packets, or  in large un-chunked data transfer scenarios, the cost of updating the control layer is much lower than the cost of data download, so resolution-based content discovery can be a good solution.

We also studied the total cost of data retrieval and observed that with limited cache storage sizes, allowing all the items to have the opportunity to be stored inside the sub-network's caches is not always the most efficient way of using the caching feature. 

For the case with a central resolution system in each sub-network and with LRU cache replacement policy, an algorithm has been proposed that can dynamically determine which items not to be cached inside the AS at any time such that the total cost of data retrieval is minimized.

In this work, our overhead model focuses on systems with Boolean states. Our future work involves systems with other state distributions. In addition, we have assumed uniform distribution of caches in the studied example. This assumption means that the probability of an item being in all the caches are the same. Future study can consider some structure like tree or power-law for the caches inside each sub-network, and using the described framework, investigate how this assumption changes the results.

\appendix{\textbf{Proof of Lemma~\ref{lem:1}}}
\begin{IEEEproof}
The distortion criteria is defined as 
\begin{eqnarray}
D_1&=&Pr(S_X=1,\hat{S}_X=0) \leq \epsilon_1 \nonumber \\
D_2&=&Pr(S_X=0,\hat{S}_X=1) \leq \epsilon_2 
\end{eqnarray}

It can be seen that $Pr(S_X=1)=\frac{\tau_X}{\tau_X+\theta_X}$, and $Pr(S_X=0)=\frac{\theta_X}{\tau_X+\theta_X}$.
There are three cases where the distortion criteria is satisfied even when the controller has no information about the underlying plane.

\begin{enumerate}
\item If the monitoring state is 'down' with high probability ($Pr(S_X=1)\leq \epsilon_1$), then having the controller assume that it is always 'down' (keeping $\hat{S}_X$ constantly equal to $'0'$) will satisfy the distortion criteria ($D_1=Pr(S_X=1)\leq \epsilon_1$ and $D_2=0<\epsilon_2$).
\item If the monitoring state is 'up' with high probability ($Pr(S_X=0)\leq \epsilon_2$), then setting the controller to assume it is always 'up' (keeping $\hat{S}_X$ constantly equal to $'1'$) will satisfy the distortion criteria ($D_1=0< \epsilon_1$ and $D_2=Pr(S_X=0)\leq \epsilon_2$).
\item If the monitoring variable can take both 'up' and 'down' states with high enough probabilities such that $1 - \frac{\epsilon_1}{Pr(S_X=1)} \leq \frac{\epsilon_2}{Pr(S_X=0)}$, then we pick a value $\rho_0$ between $1 - \frac{\epsilon_1}{Pr(S_X=1)}$ and $\frac{\epsilon_2}{Pr(S_X=0)}$, and assign $'1'$ to $\hat{S}_X$ with probability $\rho_0$ independent of the value of $S_X$. Therefore, since $D_1=Pr(S_X=1)Pr(\hat{S}_X=0)=Pr(S_X=1)(1-\rho_0) \leq \epsilon_1$, and $D_2=Pr(S_X=0)Pr(\hat{S}_X=1)=\rho_0Pr(S_X=0) \leq \epsilon_2$, the distortion criteria is satisfied.
\end{enumerate}
 
Thus in the following, we concentrate on the cases where $Pr(S_X=1)>\epsilon_1$, $Pr(S_X=0)>\epsilon_2$, and $1 - \frac{\epsilon_1}{Pr(S_X=1)}>\frac{\epsilon_2}{Pr(S_X=0)}$. 

Note that we assume that $\epsilon_1+\epsilon_2\leq 1$, then $\frac{\epsilon_2}{1-\epsilon_2}\leq \frac{1-\epsilon_1}{\epsilon_1}$, and the first two regions can be summarized in the region where $\frac{\epsilon_2}{1-\epsilon_2}\leq \frac{\theta_X}{\tau_X} \leq \frac{1-\epsilon_1}{\epsilon_1}$. The third region is also mapped to the region where $\epsilon_2 \tau_X+\epsilon_1 \theta_X < \frac{\tau_X \theta_X}{\tau_X+\theta_X}$.

Let $U^{1}_X(\epsilon_1)$ (and  $U^{2}_X(\epsilon_2)$) denote the needed update rate per change type I (and II), or in other words the ratio of times that type I (and II) changes have to be reported to the control plane so that the distortion criteria is satisfied. As can be seen in figure \ref{fig:statetime}, each 'up' period $Z_m$ starts at time $T_{2m-1}$ and ends at time $T_{2m}$. The false negative alarm is generated during the $m^{th}$ 'up' period ($Z_m$) if a type I change in the state of $X$ at time $T_{2m-1}$ is not announced to the control plane while the previous state ('0') was correctly perceived by the control plane; we show this event by $W^1_m$, and its probability is given by
\begin{eqnarray}
&Pr(W^1_m) = (1-U^1_X(\epsilon_1))Pr(\hat{S}_X=0|S_X=0)& \nonumber \\
&= (1-U^1_X(\epsilon_1))(1-Pr(\hat{S}_X=1|S_X=0))& \nonumber \\
&= (1-U^1_X(\epsilon_1))(1-\frac{Pr(S_X=0,\hat{S}_X=1)}{Pr(S_X=0)}& \nonumber \\
&= (1-U^1_X(\epsilon_1))(1-D_2\frac{\tau_X+\theta_X}{\theta_X})&  
\end{eqnarray}
 In this case, $\hat{S}_X=0$ during the time where $S_X=1$. So assuming that the $m^{th}$ such change is perceived wrong by the control plane, $Z_m$ is the time interval where the control plane has the type I wrong information about the state of $X$. Let $N_w$ be the number of times $S_X$ undergoes type I changes during a time interval $[0,w]$. The probability of type I error, and consequently type I distortion can be calculated as the ratio of total time of type I error over $w$ when $w\rightarrow \infty$.
\begin{eqnarray}
   D_1&=&E[\frac{1}{w}\sum_{m=1}^{N_w}1_{[W^1_m]}Z_m] \nonumber \\
	&=&\frac{1}{w}E[1_{[W^1_m]}Z_m]E[N_w] \nonumber \\
	&=&\frac{\tau_X}{\tau_X+\theta_X}Pr(W^1_m) \nonumber \\
	&=&\frac{\tau_X}{\tau_X+\theta_X}(1-U^1_X(\epsilon_1))(1-D_2\frac{\tau_X+\theta_X}{\theta_X})
\end{eqnarray}

Similarly, a false positive alarm is generated when a type II change is not announced while the previous perceived state ('1') was correct, and assuming that this is the $m^{th}$ such change, $Y_{m+1}$ is the time interval that the control plane has type II wrong information about $X$; let $W^2_m$ denote this event. Thus,
\begin{eqnarray}
&Pr(W^2_m) = (1-U^2_X(\epsilon_2))Pr(\hat{S}_X=1|S_X=1)& \nonumber \\
&= (1-U^2_X(\epsilon_2))\frac{Pr(S_X=1)-Pr(S_X=1,\hat{S}_X=0)}{Pr(S_X=1)} &\nonumber \\
&= (1-U^2_X(\epsilon_2))(1-D_1\frac{\tau_X+\theta_X}{\tau_X})&  
\end{eqnarray}

and 
\begin{eqnarray}
   D_2&=&E[\frac{1}{w}\sum_{m=1}^{N_w}1_{[W^2_m]}Z_m] \nonumber \\
	&=&\frac{1}{w}E[1_{[W^2_m]}Y_{m+1}]E[N_w] \nonumber \\
	&=&\frac{\theta_X}{\tau_X+\theta_X}Pr(W^2_m) \nonumber \\
	&=&\frac{\theta_X}{\tau_X+\theta_X}(1-U^2_X(\epsilon_2))(1-D_1\frac{\tau_X+\theta_X}{\tau_X})
\end{eqnarray}

To satisfy the distortion criteria we need $D_1\leq \epsilon_1$ and $D_2 \leq \epsilon_2$. The update rates per changes type I and II, $U^1_X(\epsilon_1)$ and $U^2_X(\epsilon_2)$, then can be written as

\begin{eqnarray}
&U^{1}_X(\epsilon_1) = 1-\frac{D_1 \frac{\theta_X}{\tau_X}}{\frac{\theta_X}{\tau_X+\theta_X}-D_2} \geq 1-\frac{\epsilon_1 \frac{\theta_X}{\tau_X}}{\frac{\theta_X}{\tau_X+\theta_X}-\epsilon_2}& \label{eq:U1} \\ 
&U^{2}_X(\epsilon_2) = 1-\frac{D_2\frac{\tau_X}{\theta_X}}{\frac{\tau_X}{\tau_X+\theta_X}-D_1} \geq 1-\frac{\epsilon_2\frac{\tau_X}{\theta_X}}{\frac{\tau_X}{\tau_X+\theta_X}-\epsilon_1}& \label{eq:U2}  
\end{eqnarray}
It can easily be verified that using the lower bounds obtained in equations (\ref{eq:U1}) and (\ref{eq:U2}) for update rates per each change type will result in distortions $D_1=\epsilon_1$ and $D_2=\epsilon_2$, and thus they are the minimum values needed.

Therefore, the total number of updates announced to the control plane divided by the total number of changes is given by
\begin{eqnarray}
U_X(\epsilon_1,\epsilon_2)& =& U^{1}_X(\epsilon_1) + U^{2}_X(\epsilon_2)  \label{eq:UX} 
\end{eqnarray}
Note that the total rate of type I changes, which is equal to the rate of type II changes in average is given by $\frac{1}{\tau_X+\theta_X}$ changes per second, thus total number of updates per second is given by
\begin{eqnarray}
R_X(\epsilon_1,\epsilon_2) = \frac{U_X(\epsilon_1,\epsilon_2)}{\tau_X+\theta_X} \label{eq:RX} 
\end{eqnarray}
Combining equations (\ref{eq:U1}-\ref{eq:RX}) proves the Lemma.
\end{IEEEproof}

\textbf{Proof of Lemma~\ref{lem:5}}
\begin{IEEEproof}
Recall that $\mathcal{V}_c$ is the set of caches, $\rho_i$ denotes the probability that a specific cache contains item $i$. Let $S_{ij}$ represent the state of an item $i$ at a node $j$, which is $1$ if cache $j$ contains item $i$, and $0$ otherwise, and let $\hat{S}_{ij}$ denote the corresponding state perceived by the CRS. A request from a user is not served internally (by a cache in second layer) either if no cache contains it: \begin{equation} Pr(\forall\ j\in \mathcal{V}_c:S_{ij}=0)=(1-\rho_i)^{N_c}, \end{equation}
or if there are some caches containing it but the CRS is not aware of that:

\begin{eqnarray}
  &&Pr(\exists\ j \in \mathcal{V}_c:\ S_{ij}=1\ \&\ \hat{S}_{ij}=0) \nonumber \\
  &&=\sum_{k=1}^{N_c} \sum_{1\leq j_1<..<j_k\leq N_c} Pr(^{i \notin \mathcal{V}_c-\{j_1,...,j_k\}\ \&\ }_{[\hat{S}_{i{j_l}}=0,\ S_{i{j_l}}=1]^k_{l=1}}) \nonumber \\
  &&=\sum_{k=1}^{N_c} \sum_{1\leq j_1<..<j_k\leq N_c} (1-\rho_i)^{N_c-k} \Pi_{l=1}^k Pr(^{\hat{S}_{i{j_l}}=0}_{S_{i{j_l}}=1}) \nonumber \\
  &&=\sum_{k=1}^{N_c} (^{N_c}_k) (1-\rho_i)^{N_c-k} D_{1_i}^k \nonumber \\
  &&=(1-\rho_i+D_{1_i})^{N_c}-(1-\rho_i)^{N_c} 
\end{eqnarray}

where $D_{1_i}\geq 0$ is the probability that $i$ exists in cache $j$ and the CRS does not know about it.

Thus the probability that a request is served externally is $1-P_i$ which equals
\begin{eqnarray}
&(1-\rho_i)^{N_c}+[(1-\rho_i+D_{1_i})^{N_c}-(1-\rho_i)^{N_c}]& \nonumber \\
&= (1-\rho_i+D_{1_i})^{N_c}& \label{eq:Pext}
\end{eqnarray}

where under the independent cache assumption, the state of an item in a cache is independent of the state in another cache. The probability $D_{1_i}\geq 0$ is always less than the probability of $i$ being in cache $j$ ($D_{1_i}\leq \rho_i$), and if the state updates are done at rate greater than $R_i(\epsilon_1,\epsilon_2)$, it will also be less than $\epsilon_1$. 
\end{IEEEproof}

\bibliographystyle{IEEEtran}
\vspace{-0.6em}
\bibliography{IEEEabrv,TCOM-TPS-16-1006_R1}

\end{document}